\input{aipcheck}

\documentclass[
    ,final            
  ]
  {aipproc}

\layoutstyle{6x9}

\begin{document}

\title{{\it XMM--Newton}\, EPIC \& OM Observations of Her~X-1 over the 
35~d Beat Period and an Anomalous Low State}

\author{S. Zane}{address={Mullard Space Science Laboratory, 
University College of London, 
Holmbury St Mary, 
Dorking, Surrey, RH5 6NT,  UK
}, email={sz@mssl.ucl.ac.uk}}

\author{G. Ramsay}{address={
Mullard Space Science Laboratory, 
University College of London, 
Holmbury St Mary, 
Dorking, Surrey, RH5 6NT,  UK
}, email={gtbr@mssl.ucl.ac.uk}}

\author{Mario A. Jimenez-Garate}{address={
MIT Center for Space Research, 77 Massachusetts Avenue, Cambridge, MA 
02139, USA}}

\author{Jan Willem den Herder}{address={
SRON, the National Institute for Space Research,
Sorbonnelaan 2,  3584 CA Utrecht, The Netherlands}} 

\author{Martin Still}{address={NASA/Goddard Space Flight Center, Code 662, 
Greenbelt, MD 20771}} 

\author{Patricia T. Boyd}{address={NASA/Goddard Space Flight Center, Code 
662, Greenbelt, MD 20771}}

\author{Charles J. Hailey}{address={Columbia Astrophysics Laboratory, 
Columbia University, New York, NY 10027, USA}}

\begin{abstract}
We present the results of a series of {\it XMM--Newton}\, EPIC and OM 
observations of
Her~X-1, spread over a wide range of the 35 d precession period. We
confirm that the spin modulation of the neutron star is weak or absent in
the low state - in marked contrast to the main or short-on states. The
strong fluorescence emission line at $\sim 6.4$~keV is detected in all
observations (apart from one taken in the middle of eclipse), with higher
line energy, width and normalisation during the main-on state. In
addition, we report the detection of a second line near 7 keV in 10 of the
15 observations taken during the low-intensity states of the system.
We discuss these observations in the context of previous observations, 
investigate the origin of the soft and hard X-rays and consider the 
emission site of the 6.4keV and 7keV emission lines.
\end{abstract}

\maketitle


\section{Introduction}

Her X-1 is one of the best studied X-ray binaries in the sky. The binary
system consists of a neutron star and an A/F secondary star. It has an
orbital period of 1.7~d and the neutron star spin period is $\sim$1.24~s.
It is one of only a few systems which shows a regular variation in X-rays,
over a ``beat'' period of 35~d, which is generally interpreted as the
precession of an accretion disk that periodically obscures the neutron
star beam. The cycle comprises: i) a ~10 d duration main-on state, ii) a
fainter ~5 d duration short-on state, and iii) a period of lower emission
in between.  Exceptions to the normal 35 d cycle has been observed only in
four occasions: in 1983, 1993, 1999 and in January 2004 the turn-on of the
source has not been observed.  During these "anomalous low states" 
(ALS; the period of which ranges from several months to 1.5 years) Her X-1 
appears as a
relatively faint X-ray source, with a strength comparable to that of the
standard low state. The event registered last year was only the fourth one 
that has been seen since the
discovery of Her X-1 in 1972, and the first one that could be
observed with a high capability satellite as {\it XMM--Newton}. 

Her X-1 has been observed by {\it XMM--Newton}\, on 15 separate 
occasions outside the ALS, 
giving good coverage over the beat period. Moreover, 
it has been observed 10 times during the ALS. The analysis of datasets 
taken before the ALS has been presented 
by Ref.~\cite{r02}, \cite{mario02} and \cite{io04}. 
Here we report our main findings, referring the interested reader to the 
above papers for more details, and we present a preliminary analysis of 
the ALS. 

\section{Timing analysis}
\label{sec:tim}
We first focussed on datasets taken before the source entered the 
anomalous low state. 
We performed a search for pulsations in all datasets, confirming that, 
in marked contrast to the main or short-on states, 
the spin modulation of the neutron star is weak or absent in the low
state. During the states of higher intensity, we observe a
substructure in the broad soft X-ray modulation below $\sim 1$~keV, 
revealing the presence of separate peaks which reflect the structure seen
at higher energies (see Ref.~\cite{io04}). 

\begin{figure}
\setlength{\unitlength}{1cm}
\begin{picture}(8,3.8)
\put(.2,-.3){\includegraphics{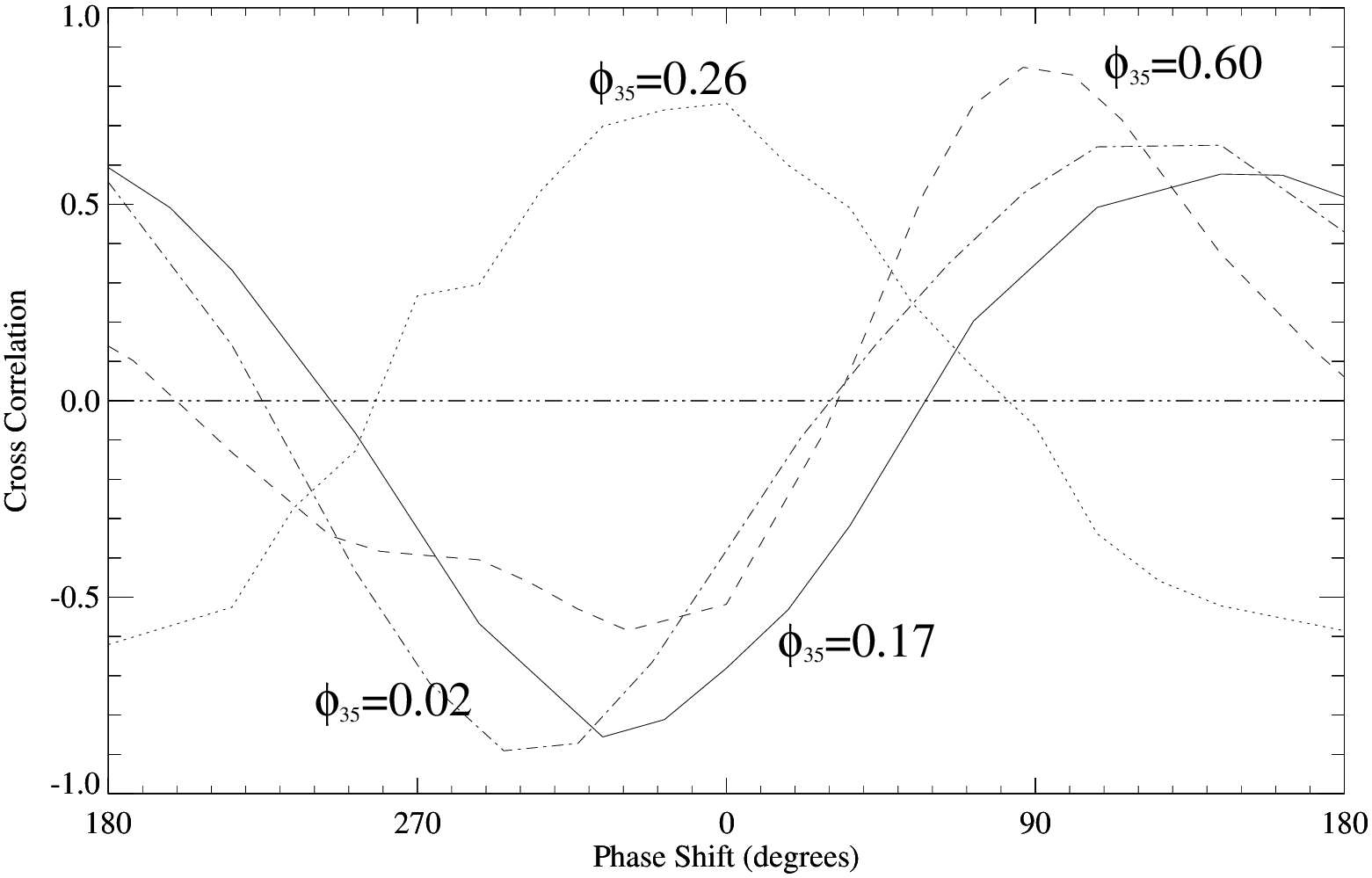}}
\end{picture}
\caption{Cross correlation of the (0.3-0.7~keV) and
(2-10~keV) light curves at four different $\Phi_{35}$.
\label{cross_cor}}
\end{figure}

The soft and hard X-ray lightcurves of Her X-1 are known to be shifted in
phase (see Fig.\ref{6.4line}, central panel). 
Under the assumption that soft X-rays
are due to the reprocessing of the pulsar beam by the inner edge of
the disk, this is usually interpreted as evidence for a tilt angle in
the disk\cite{oo97}$, $\cite{oo00}. In our {\it XMM--Newton}\, 
data, we find the first evidence
for a substantial and systematic change in the phase difference along the 
beat cycle, which is predicted by precessing disk models\cite{gb76} (see
Fig.~\ref{cross_cor}).

\section{The Fe K$\alpha$ line}

\begin{figure}
\setlength{\unitlength}{1cm}
\begin{picture}(8,12.3)
\put(-6.8,11.7){\includegraphics{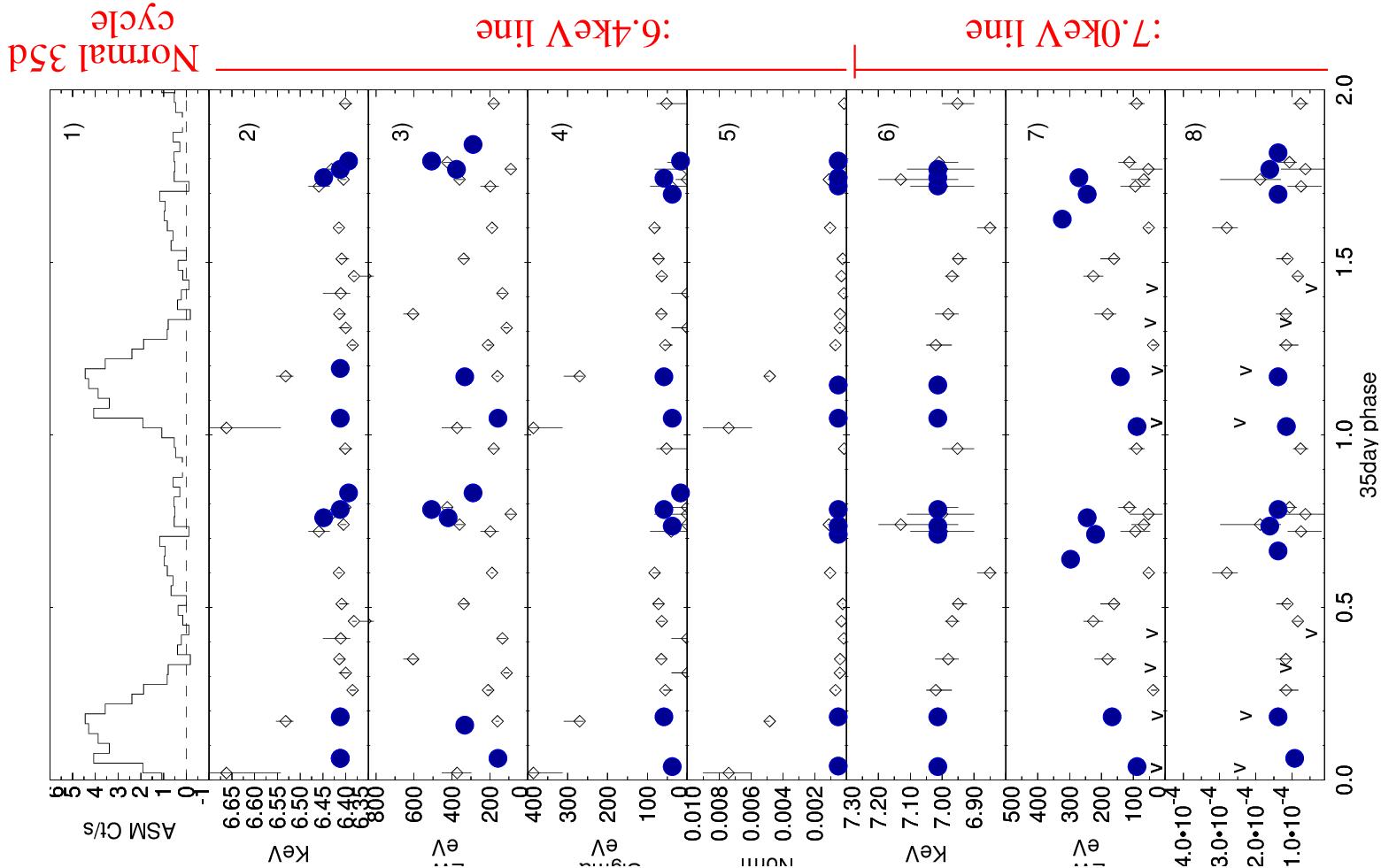}}
\put(1.2,-.2){\includegraphics{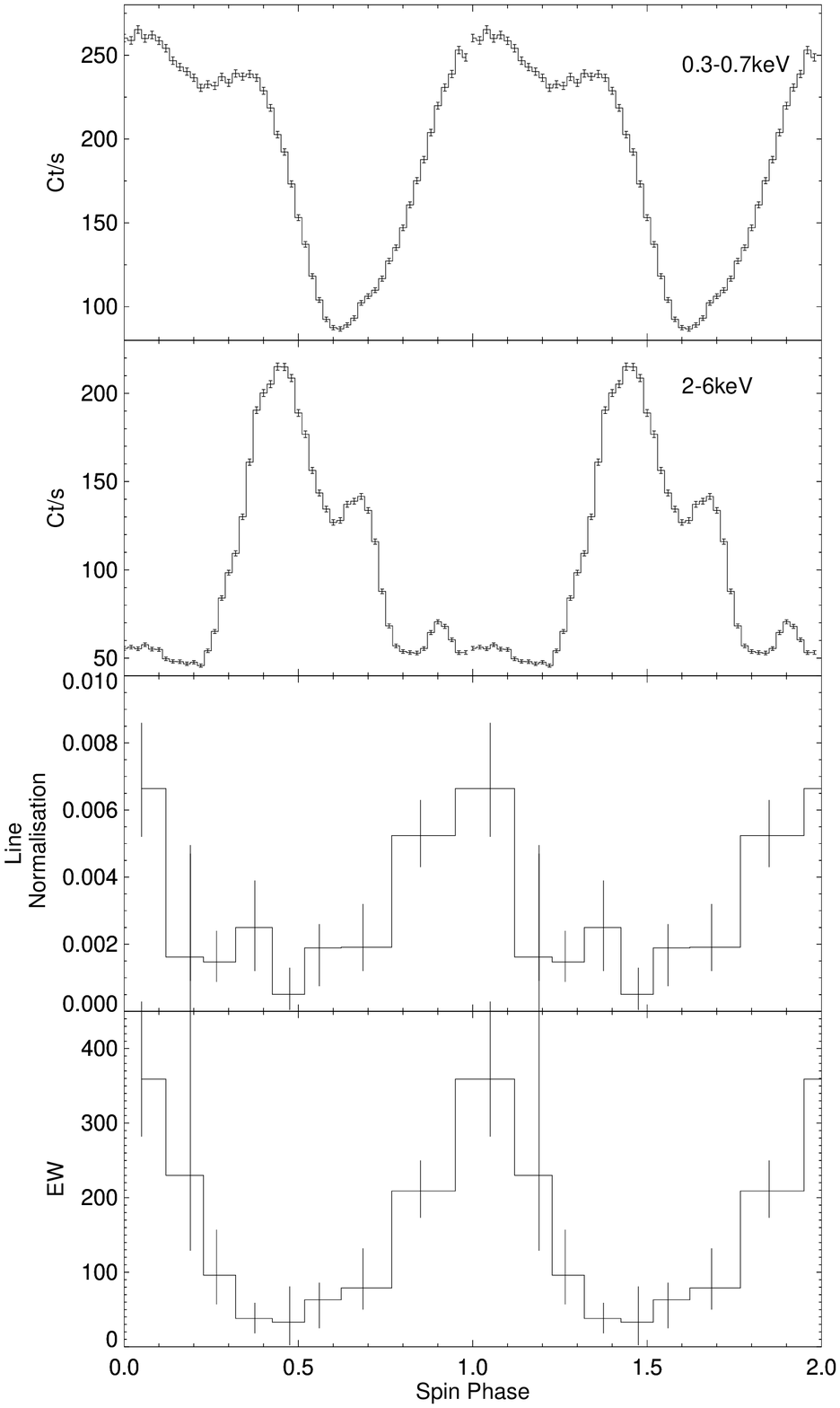}}
\put(5.3,11.5){\includegraphics{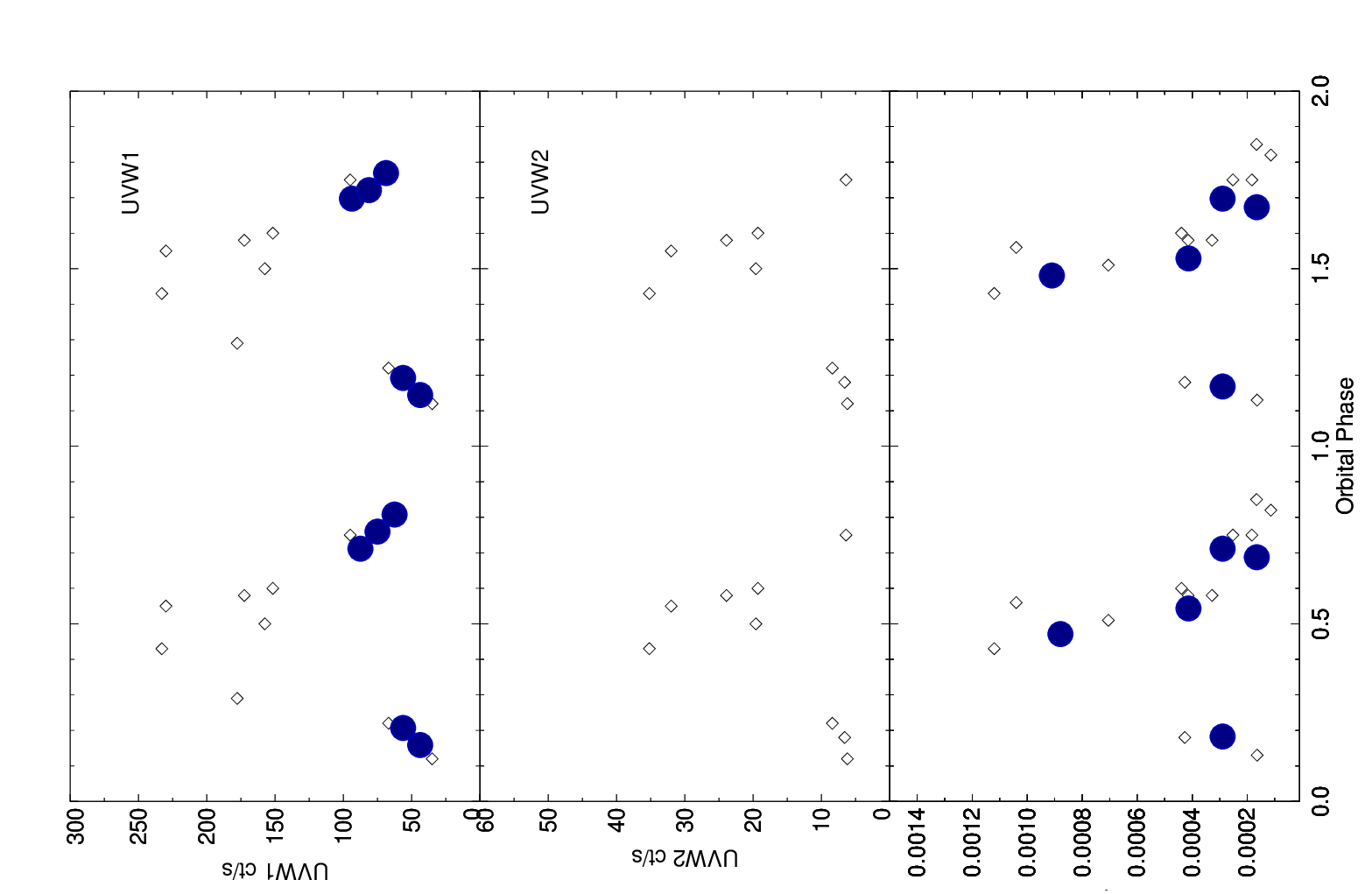}}
\end{picture}
\caption{Left panel:
Variation of the K$\alpha$ and Fe~XXVI line parameters along
the beat cycle. 
From the top: 1) mean ASM light curve; 2)-5) central
energy, equivalent width (EW), width and normalization of the
prominent K$\alpha$ Fe emission line; 6)-8) central energy, EW and 
normalization of the Fe line at $\sim 7$~keV. When 
the second feature is undetected, we show the 90\% confidence level upper 
limits to the EW and normalization (``v'' symbols).
Central panel: The observation made at $\Phi_{35}$=0.02. From the top 
panel: the light curve in the 0.3-0.7~keV and 2-6~keV band (note the 
shift in phase between hard and soft X-ray emission); the Fe
K$\alpha$ line normalisation and EW as a function of the spin
phase. Right:
The UVW1 (top panel), UVW2 (central panel) data and the Fe 6.4~keV
normalisation folded on the orbital period. The blue circles 
superimposed in the left and right panels are data taken 
during the 2004 ALS. }
\label{6.4line}
\end{figure}

The strong emission line at $\sim 6.4$~keV is detected in all our {\it 
XMM--Newton}\, 
observations, with larger broadening and normalization during the
main-on (see Fig.~\ref{6.4line}, left panel). 
The line centroids observed using the EPIC PN deviate by $4 \sigma$
from the 6.40~keV neutral value: the Fe line emission
originates in
near neutral Fe (Fe XIV or colder) in the low and short-on state
observations, whereas in the main-on the observed Fe K$\alpha$ centroid
energies ($6.65 \pm 0.1$~keV and $6.50 \pm 0.02$~keV
at $\Phi_{35}=0.02$ and 0.17) correspond to Fe XX-Fe XXI \cite{pa03}. 

Possible reasons for this behaviour may be: 1) an array of Fe
K$\alpha$ fluorescence lines exists for a variety of charge states of Fe
(anything up to Fe XXIII); 2)
Comptonization from a hot corona for a narrower range of charge states
centered around Fe XX; 3) Keplerian motion. The Keplerian velocity 
measured at $\Phi_{35}=0.02$ and 0.17 is $\sim 15500$ and $\sim 13000$ 
km/s,
respectively. This gives a radial distance of $\sim 2-3 \times
10^8$~cm, which is close to
the magnetospheric radius for a magnetic field of $\sim 10^{12}$~G.

However, another possibility is that the region
responsible for the Fe K$\alpha$ line emission is different
for lines observed at different beat phases. In fact, while data taken 
during the main-on clearly indicate a correlation between the fluorescent 
Fe K$\alpha$
line and the soft X-ray emission (Fig.~\ref{6.4line}, central panel),  
suggesting a common origin in the illuminated hot spot at the inner edge 
of the disk, the same is not explicitly evident in
data taken during the low state. Instead, at such phases the  Fe K$\alpha$ 
line is a
factor $>5$ weaker and is clearly modulated with the orbital period
(see again Fig.~\ref{6.4line}, right panel). The correlation between the
fast rising UVW1 flux and the Fe K$\alpha$ detected outside the main-on  
point to a common origin, possibly in the disk and/or 
illuminated companion.

A fraction of the Fe K$\alpha$  emission may arise from
relatively cold material in a disk wind, such as commonly observed in
cataclysmic
variables \cite{drew97}. However, we do not detect
the Fe K$\alpha$
line during the middle of the eclipse, and the
upper limit on the line flux is
$\sim10$ or less of that measured outside the eclipse. Also, there is no
Doppler signature of a wind in the HETG spectrum of the Fe K$\alpha$
line \cite{mario05}. On the other hand, the data reported here suggest a
complex origin for the overall emission of the Fe K$\alpha$ line. To our
knowledge, a complex of lines which include all ionization states from Fe
XVIII to XXIV Fe K$\alpha$ has not been observed in any astrophysical
source. This may still indicate an outflow of relatively cold gas or some
complex dynamics in the disk/magnetosphere interface.
Such phenomena
should be time-dependent and may be monitored in the future using
Astro-E2.

\section{The $\sim7$~keV Fe line}
\label{7keV}

\begin{figure}
\setlength{\unitlength}{1cm}
\begin{picture}(8,4.5)
\put(-3.2, -.2){\includegraphics{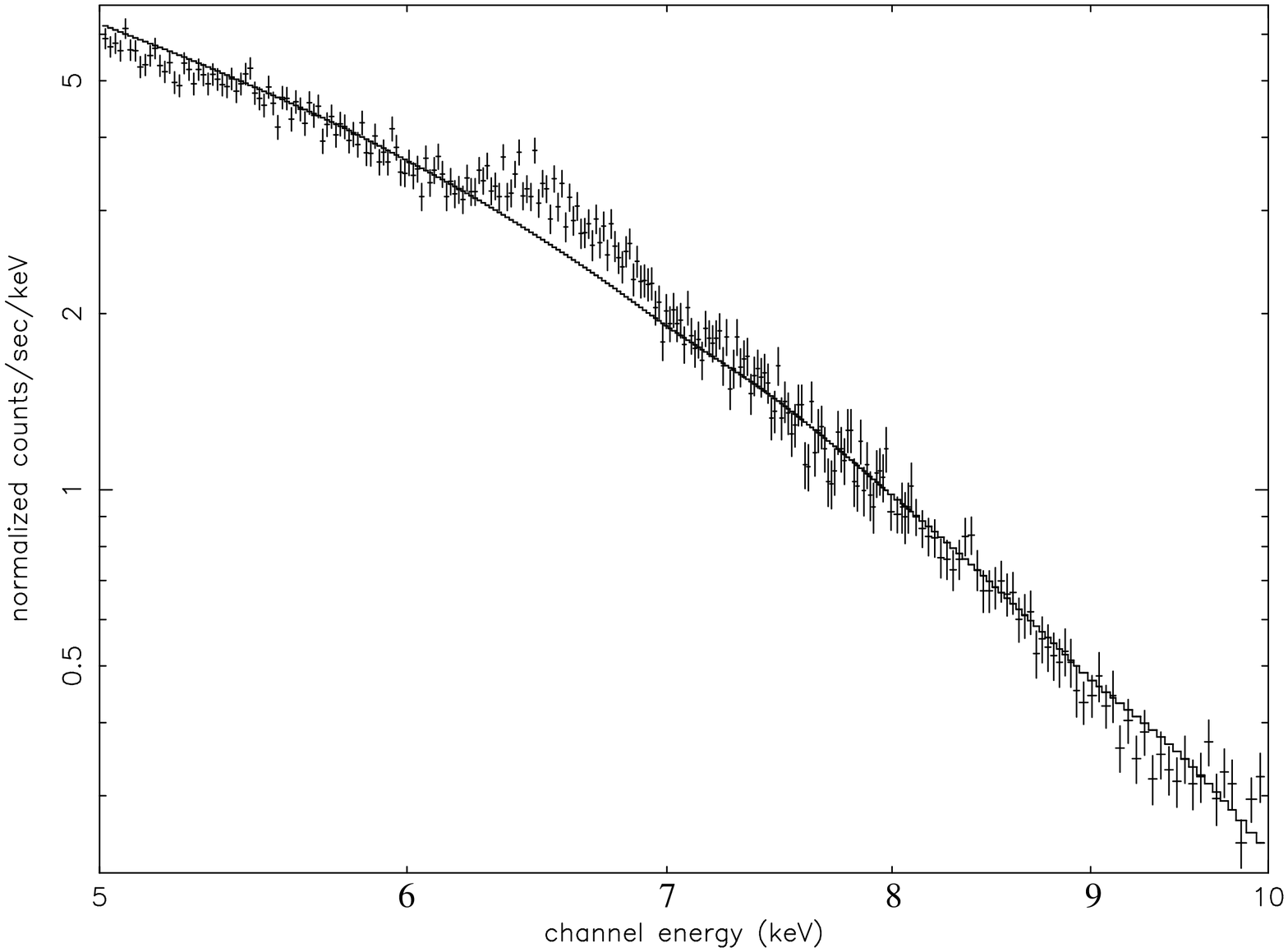}}
\put(4.2, -.2){\includegraphics{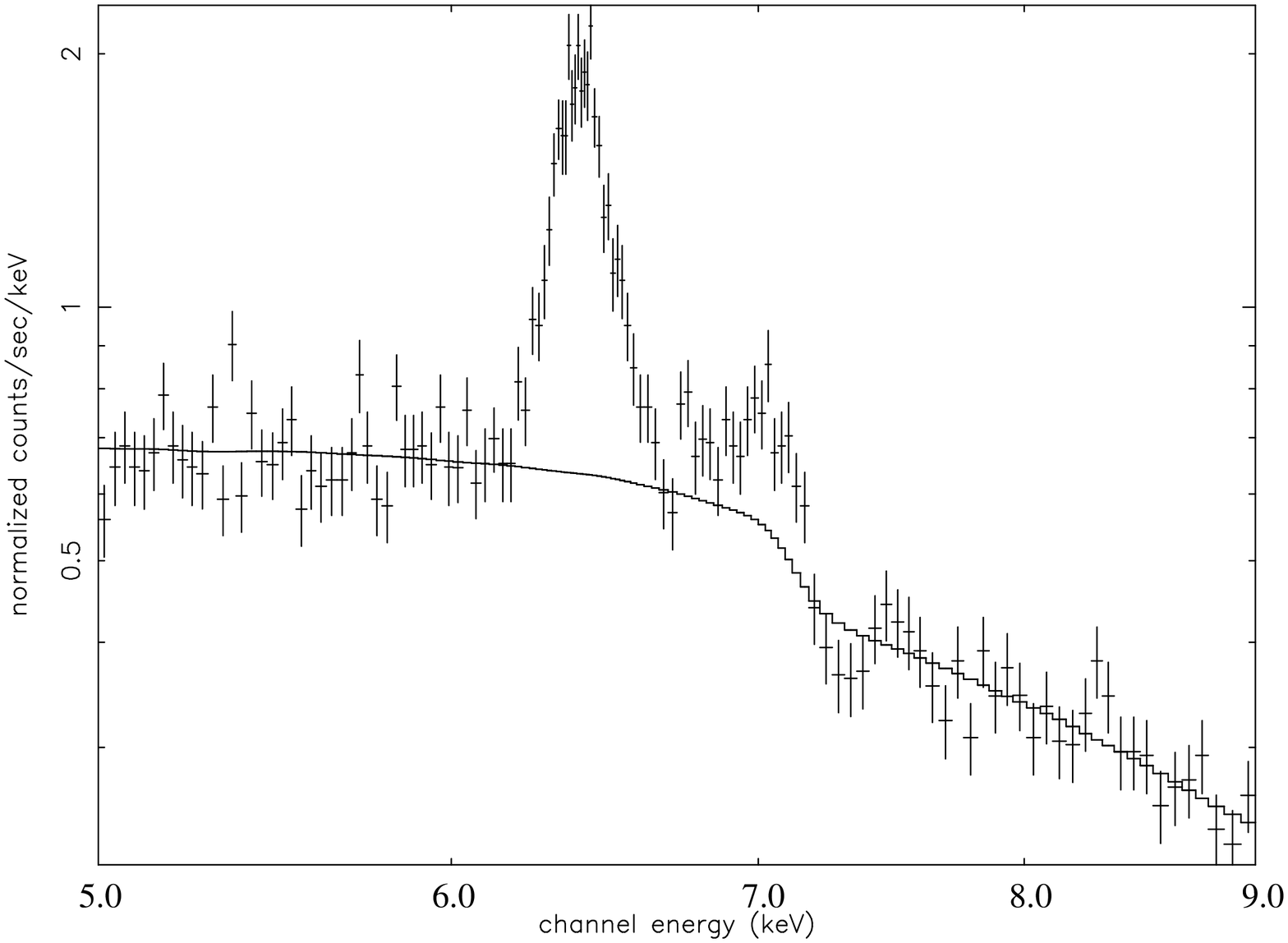}}
\end{picture}
\caption{The spectral region around 6.6keV from $\Phi_{35}$=0.02 (top) and
$\Phi_{35}$=0.79 (bottom). In both panel a solid line
shows the best fit after the normalisation of the one or two Gaussian
components have been set to zero.} \label{fe_zoom}
\end{figure}

{\it XMM--Newton}\, data have revealed, for the very first time for this 
source, the presence of a second Fe line at $\sim7$~keV. The 
feature is only detected during the low and short-on states, and over 
several beat phases (see Fig.~ \ref{fe_zoom}). Also, it has been 
confirmed by a {\sl Chandra} HETGS observation of the source (the only one 
made during the low state) taken at $\Phi_{35}= 0.44-0.46$ \cite{mario05}. 

The feature cannot be produced by fluorescence, and it is more likely to 
be a Fe~XXVI line originating in widely extended photo-ionized plasma.
This is consistent with the fact that also 
RGS data taken during the low and short-on states 
show the presence of photo-ionized gas \cite{mario02}. 
Grating spectra exhibit several narrow recombination emission lines,
the most prominent being C~VI, N~VI, N~VII, O~VII, O~VIII and
Ne~IX.
The line ratio $G=(f+i)/r$, as computed for all the helium-like
ion complexes, is $G \simeq 4$, which indicates that photoionization
is the dominant mechanism. Moreover, RGS spectra shows two radiative
recombination continua of O~VII and N~VII, consistent with a low
temperature of the emitting plasma
(30000~K~$<T<$~60000~K) \cite{mario02}. 
None of these features is detected during the
main-on state. 

The recombination X-ray line emission are not likely to 
originate in HZ Her, due to the absence of UV induced photoexcitation
signatures in the He-like triplets (observed with HETGS) \cite{mario05}. 
Instead, we propose that an extended,
photoionized accretion disk atmosphere may be responsable
for such features \cite{mario02}. The evidence for the disk
identification relies on the modeled structure and spectra from a
photoionized disk, which is in
agreement with the limit set spectroscopically on the density of the 
low-energy
lines emitting region. This theoretical model has been developed by 
Ref.~\cite{mario01}, who computed the spectra
of the atmospheric layers of a Shakura-Sunyaev accretion disk,
illuminated by a central X-ray continuum. They found that, under
these conditions, the disk develops both an extended corona which is
kept hot at the Compton temperature, and a more compact, colder, X-ray
recombination-emitting atmospheric layer (see Figure~9 in
Ref.~\cite{mario02}).

Interestingly, we find that the Fe XXVI line detected by {\it 
XMM--Newton}\, may be a
signature of the {\it hottest} external layers of the disk corona,
which are located above the recombination-emitting layers.
Again, the computed values of the density agree with the constraints
inferred from the 7~keV line parameters \cite{io04}.

In summary, most of the spectral lines 
discovered with EPIC and RGS can be associated with the illuminated
atmosphere/corona of the accretion disk, which explains why they are more 
prominent during the low states when the direct X-ray beam from the pulsar 
is obscured by the accretion disk. Therefore, the variability of the 
Her~X-1 spectrum lends support to the precession of the accretion disk,
strengthen the interpretation of the low state emission in term of an
extended source and open the exciting possibility to monitor
spectroscopically the different atmospheric components of the disk
during the transition from the low to the high state.

\section{The Anomalous Low State}

Anomalous low states are rare and peculiar events, during which the source 
resides in a deep low state. While the 
mechanism that forces state changes is almost certainly variations in 
accretion disk structure, the engine ultimately driving structural 
evolution remains unknown. 
Each past anomalous low state, including the most recent one of January 
2004\cite{mart04}, has been preceded by a period of enhanced
spin-down, that has been interpreted in terms of an increasing torque
leading to a reversal in the rotation of the inner disk. This also implies
that the onset of an anomalous low state is accompanied by a large
variation in the structure of the inner region of the accretion disk, that
becomes increasingly warped.  

In order to search for residual evidence of a 35 d cycle, we compared the 
line emission detected during the ALS with that observed in the several 
{\it XMM--Newton}\ datasets we have accumulated during the standard 35 d 
cycle. As we can see from Fig.~\ref{6.4line}, as far as the Fe complex is 
concerned, there is no 
an obvious difference in the spectral properties of the ALS and of the 
standard low states. The line features are consistent with being 
the same in these epoches and the orbital variation of the K$\alpha$ Fe 
line shows the same  correlation with the UVW1. This supports  
our scenario in which the 
Fe line emission of the low state originates in an extended component 
(disk atmosphere/corona and/or companion) instead that in the inner region 
of the disk. Moreover, it is consistent with the fact that at higher 
energies a significant 
Compton reflection component has been detected by RXTE in the spectrum of 
the ALS (\cite{marthead}). 

A detailed comparison of the recombination emission lines measured by RGS 
can shed more light on this issue. If our interpretation is correct, by 
measuring the line ratios during the anomalous low state allows to infer 
the ionization state of the plasma, column density and optical depth in the 
visible portion of the disk
atmosphere/corona, ultimately  constraining the accretion geometry.


\bibliographystyle{aipproc}   

\bibliography{sample}

\IfFileExists{\jobname.bbl}{}
 {\typeout{}
  \typeout{******************************************}
  \typeout{** Please run "bibtex \jobname" to optain}
  \typeout{** the bibliography and then re-run LaTeX}
  \typeout{** twice to fix the references!}
  \typeout{******************************************}
  \typeout{}
 }

\end{document}